\documentclass[prb,twocolumn,superscriptaddress]{revtex4}

\usepackage{graphicx}
\usepackage{dcolumn}
\usepackage{amsmath}
\usepackage{color}

\begin{document}


\title{Effect of doping on the electronic structure, orbital-dependent renormalizations, and magnetic correlations in bilayer La$_3$Ni$_2$O$_7$}  

\author{I. V. Leonov}
\affiliation{M. N. Mikheev Institute of Metal Physics, Russian Academy of Sciences, 620108 Yekaterinburg, Russia}
\affiliation{Institute of Physics and Technology, Ural Federal University, 620002 Yekaterinburg, Russia}

\begin{abstract}
Using the DFT+dynamical mean-field theory (DFT+DMFT) approach we study the effects of electronic correlations and doping on the normal state electronic structure of the double-layer nickelate superconductor La$_3$Ni$_2$O$_7$ under pressure. In agreement with experiments, we obtain significant orbital-dependent quasiparticle renormalizations of the Ni $x^2-y^2$ and $3z^2-r^2$ bands, accompanied by incoherence (bad metal behavior) of the $3z^2-r^2$ states, caused by the proximity of the Ni $3d$ states to orbital-dependent localization. Our results demonstrate a sensitive, non-monotonic dependence of $m^*/m$ on doping, with a remarkable, by about 20\%, increase for the Ni $x^2-y^2$ orbitals upon electron doping $x \sim 0.2$ (per Ni ion), implying a significant enhancement of orbital-dependent correlations with oxygen deficiency in LNO. We observe a reconstruction of the low-energy electronic structure of LNO upon doping above $x\sim -0.3$ and 0.2. It is associated with the Lifshitz transition, with a crossover to a self-doping regime characterized by partial occupation of the La $5d$ bands (upon an electron doping $x>0.2$). Our analysis of the static magnetic susceptibility $\chi({\bf q})$ obtained within DFT+DMFT suggests the possible formation of the spin and charge (or bond) density wave stripes, implying strong spin and charge correlations in LNO. We show that this behavor is associated with suppression of the N\'eel $G$-type antiferromagnetic ordering of the Ni$^{2+}$ ions upon hole doping. Interestingly, upon a moderate electron doping of the Ni$^{2.5+}$ ions (e.g., with oxygen deficiency), we find a significant enhancement of the strength of in-plane spin and charge fluctuations. We note a close resembles of our results to those for the bilayer Hubbard model, which shows the boosting of superconductivity as one of the two electron bands approaches the Lifshitz transition (e.g., upon doping). Our results suggest that spin and charge stripe fluctuations, effectively tuned by doping, play a key role in pressure-driven superconductivity in LNO.
\end{abstract}

\maketitle

\section{Introduction}

The recent discovery of superconductivity with a high critical temperature, up to $T_c \sim 80$-100~K, comparable to that in superconducting cuprates, in the highly-pressurized bulk and then later in the ambient-pressure thin films of the double-layer Ruddlesden-Popper (RP) nickelate La$_{n+1}$Ni$_n$O$_{3n+1-\delta}$ (LNO) with $n=2$ has generated significant experimental and theoretical interest to this novel class of superconducting materials \cite{Sun_2023,Wang_2024a,Hou_2023,YZhang_2024,
Li_2025,Qiu_2025,Ko_20205,Zhou_2025,Dong_2024,Yang_2024a,
Liu_2024,Xie_2024,MZhang_2024,Liu_2025,Chen_2024a,Kakoi_2024,
Agrestini_2024,Dan_2024,YYang_2023,Qin_2023,
Meng_2024,Khasanov_2024,Zhang_2023a,Shilenko_2023,
Lechermann_2023,Christiansson_2023,Liao_2023,Shen_2023b,
Ryee_2024,Lechermann_2024,Craco_2024,Cao_2024,Wu_2024,
Leonov_2025,LaBollita_2024b,BZhang_2024,Ni_2024,Tian_2025,
Zhang_2023c,Liu_2023b,Yang_2023,Heier_2024,
Lu_2024,Fan_2024,Sakakibara_2023a,Tian_2024a,MWang_2024,Shi_2025}. While being structurally related to the infinite-layer hole-doped nickelates $R$NiO$_2$ with $R=\mathrm{REE}$, Sr, Ca (with a nominal Ni$^+$ electronic configuration), which exhibit superconductivity below $\sim$15~K \cite{Li_2019,Hepting_2020,Osada_2021,Lee_2023}, with a modest increase to $\sim$31 K upon optimizing their crystalline quality, compositions, epitaxial strain, and pressure \cite{Wang_2022,Ren_2023}, the double-layer RP LNO (with $n=2$) adopt a different nominal electronic configuration Ni$^{2.5+}$ (Ni $3d^{7.5}$ state with an equal amount of Ni$^{2+}$ and Ni$^{3+}$ ions in the unit cell) with two Ni $e_g$ orbitals ($x^2-y^2$ and $3z^2-r^2$) contributing to the Fermi surface. This implies that the low-energy electronic structure of LNO distinct from that in the infinite-layer systems, suggesting different microscopic mechanisms may be at play to explain superconductivity in these materials.

Moreover, recent experiments show superconductivity with $T_c \sim 30$-40 K to appear in the La- and Pr-based trilayer RP nickelates (with $n=3$) under pressures above 15 GPa \cite{Zhu_2024,Li_2024,Zhang_2025,MZhang_2025,Pei_2026}. Note that in the trilayer LNO we deal with the formal oxidation state of Ni$^{2.67+}$, with two Ni$^{3+}$ and one Ni$^{2+}$ ions in the unit cell \cite{Wang_2024,Leonov_2024a,LaBollita_2024a,
Huang_2024,Chen_2024b,Tian_2024b,ZhangLin_2024b,
Yang_2024b}. It is remarkable that superconductivity in both the double-layer and trilayer LNOs is found to appear near a pressure-driven structural phase transition to the high-symmetry (orthorhombic or tetragonal) phase, which is characterized by the absence of tilting of NiO$_6$ octahedra. Moreover, recent single-crystal x-ray diffraction, transport, muon spin relaxation ($\mu^+$ SR), and nuclear magnetic resonance measurements suggest that a spin-charge-density wave stripe ordering sets in LNO upon cooling below about 150 K (at ambient pressure) \cite{Zhu_2024,Li_2024,Zhang_2025,MZhang_2025,Pei_2026,Chen_2024a,
Kakoi_2024,Agrestini_2024,Dan_2024}. This result corroborates with recent resonant inelastic x-ray diffraction and inelastic neutron diffractions measurements of magnetic excitation in LNO, which show the presence of the in-plane magnetic correlations associated with spin-charge density wave ordering \cite{Chen_2024a,Kakoi_2024,Agrestini_2024,Dan_2024}. In fact, superconductivity in the RP LNO sets near a structural phase transition to the high-symmetry phase, (most presumably) associated with suppression of a long-range spin-charge-density wave state under pressure. This suggests that spin and charge stripe fluctuations play an important role to tune superconductivity in LNO under pressure.

In spite of active research, microscopic understanding of the properties of superconducting nickelates still remains challenging. In fact, recent theoretical analysis based on applications of the state-of-the-art electronic structure techniques, e.g., such as the DFT+dynamical mean-field theory (DMFT) \cite{Georges_1996,Kotliar_2006} and GW+DMFT \cite{Biermann_2003,Tomczak_2017} methods, suggest the crucial importance of the effects of strong correlations, associated with the partially occupied Ni $3d$ shell of LNO \cite{Shilenko_2023,
Lechermann_2023,Christiansson_2023,Liao_2023,Shen_2023b,
Ryee_2024,Lechermann_2024,Craco_2024,Cao_2024,Wu_2024,
Leonov_2025}. These calculations show the complex interplay between the effects of electron correlations, Fermi surface nesting, and pressure \cite{Shilenko_2023,Lechermann_2023,
Christiansson_2023,Ryee_2024,Lechermann_2024,Liao_2023,Shen_2023b}. In addition, the effects of correlations exhibit a remarkable orbital selectivity, such as a strong orbital dependence of the quasiparticle mass renormalizations $m^*/m$ and incoherence of the spectral weights of the Ni $3d$ states in LNO. In agreement with this, experiments suggest a weakly insulating behavior at low pressures and the formation of bad (strange) metal phase above $T_c$ in LNO \cite{Sun_2023,Wang_2024a,Hou_2023,YZhang_2024,Li_2025,Qiu_2025}.

While various theoretical analyses converge towards the $s_{\pm}$-type superconductivity in LNO (possibly competing with the $d$-wave component) \cite{Zhang_2023c,Liu_2023b,Yang_2023,
Heier_2024,Lu_2024,Fan_2024,Sakakibara_2023a,Tian_2024a}, its microscopic mechanisms, as well as the origins of anomalous normal state properties of LNO still remain debating. 
In our study, using the fully self-consistent in charge density DFT+DMFT approach \cite{Georges_1996,Kotliar_2006,Leonov_2020a,Leonov_2024b} we explore the normal state electronic properties of the double-layer LNO. We study the effects of electron-electron correlations and chemical doping, which yield remarkable modifications of the electronic structure and magnetic correlations of LNO. In fact, we observe that the properties of LNO are strongly affected by the interplay between strong correlations and the precise level of doping (stoichiometry). Our DFT+DMFT results suggest the possible formation of spin and charge (or bond) density wave stripes, implying strong spin and charge correlations in LNO. We show that this behavor is associated with suppression of the long-range N\'eel $G$-type antiferromagnetic (AFM) ordering of the Ni$^{2+}$ ions upon hole doping. Interestingly, upon a moderate electron doping of the Ni$^{2.5+}$ ions (e.g., upon oxygen deficiency) we find a significant enhancement of the in-plane spin and charge fluctuations. Our results are consistent with the behavior of the bilayer Hubbard model, which shows enhancement of superconductivity as one of the two electron bands approaches a Lifshitz transition and, in particular, when it becomes incipient (e.g., due to doping) \cite{Karakuzu_2021}. We show that this behavior associated with a significant enhancement of the strength of in-plane spin and charge fluctuations in LNO. 


\section{Computational details}

In this work, using DFT+DMFT we explore the effects of electron-electron correlations and chemical doping on the electronic structure, orbital-dependent quasiparticle renormalizations, Fermi surface topology, and magnetic correlations of the high-pressure phase of the double-layer LNO. In our calculations we employ the fully self-consistent in charge density DFT+DMFT method implemented with plane-wave pseudopotentials \cite{Leonov_2020a,Leonov_2024b}. In DFT we use generalized gradient approximation with the Perdew-Burke-Ernzerhof exchange functional as implemented in the Quantum ESPRESSO package \cite{Giannozzi_2009, Giannozzi_2017, DalCorso_2014}. As a start, we perform structural optimization of internal atomic positions of LNO for the experimentally determined at about 30 GPa crystal structure (space group $Fmmm$) within non-magnetic DFT.
 To model the effects of chemical doping on the electronic structure of LNO, we employ a rigid-band shift of the Fermi level within DFT taking into account the effects of charge transfer and strong correlations of the Ni $3d$ states within the fully self-consistent in charge density DFT+DMFT approach \cite{Leonov_2020a,Leonov_2024b}. In our DFT+DMFT calculations we explicitly include the Ni $3d$, La $5d$, and O $2p$ valence states, by constructing a basis set of atomic-centered Wannier functions within the energy window spanned by these bands \cite{Anisimov_2005,Marzari_2012}. This allows us to take into account a charge transfer between the partially occupied Ni $3d$, La $5d$, and O $2p$ states, accompanied by the strong electron-electron interactions in the Ni $3d$ shell.

We employ the continuous-time hybridization expansion (segment) quantum Monte Carlo algorithm in order to solve the realistic many-body problem \cite{Gull_2011}. The effects of electron correlations in the Ni $3d$ shell are treated by using the on-site Hubbard $U = 6$ eV and Hund's exchange $J = 0.95$ eV, taken in accordance with previous DFT+DMFT calculations of nickelates. The La $5d$ and O $2p$ valence states are uncorrelated and are treated on the DFT level within the self-consistent DFT+DMFT approach. In our calculations we use the fully localized double-counting correction evaluated from the self-consistently determined local occupations. We neglect the spin-orbit coupling in the DFT+DMFT calculations for simplicity (which is expected to be small for the Ni ion). Using Pad\'e approximants we perform analytic continuation of the self-energy results from the Matsubara frequencies [$\Sigma(i\omega_n)$] to the real energy axis [$\Sigma(\omega)$] in order to compute the {\bf k}-resolved spectra and quasiparticle Fermi surfaces.

Using DFT+DMFT we study the normal-state electronic properties of the paramagnetic (PM) phase of double-layer LNO at a temperature $T=116$~K. In this work, we discuss the impact of doping on the electronic structure of LNO (near to its stoichiometric composition it has a nominal Ni$^{2.5+}$ electron configuration). The doping was tuned from the Ni$^{2+}$ (electron doping $x=0.5$ per Ni ion) to Ni$^{3+}$ ion (hole doping $x=-0.5$), neglecting by the effects of structural transformations and vacancy formations driven by doping. The latter allows us to trace out the changes of the electronic structure, magnetic state, and Fermi surfaces of highly-pressurized LNO.

\section{Results and discussion}
\subsection{Electronic structure}

\begin{figure}[tbp!]
\centerline{\includegraphics[width=0.5\textwidth,clip=true]{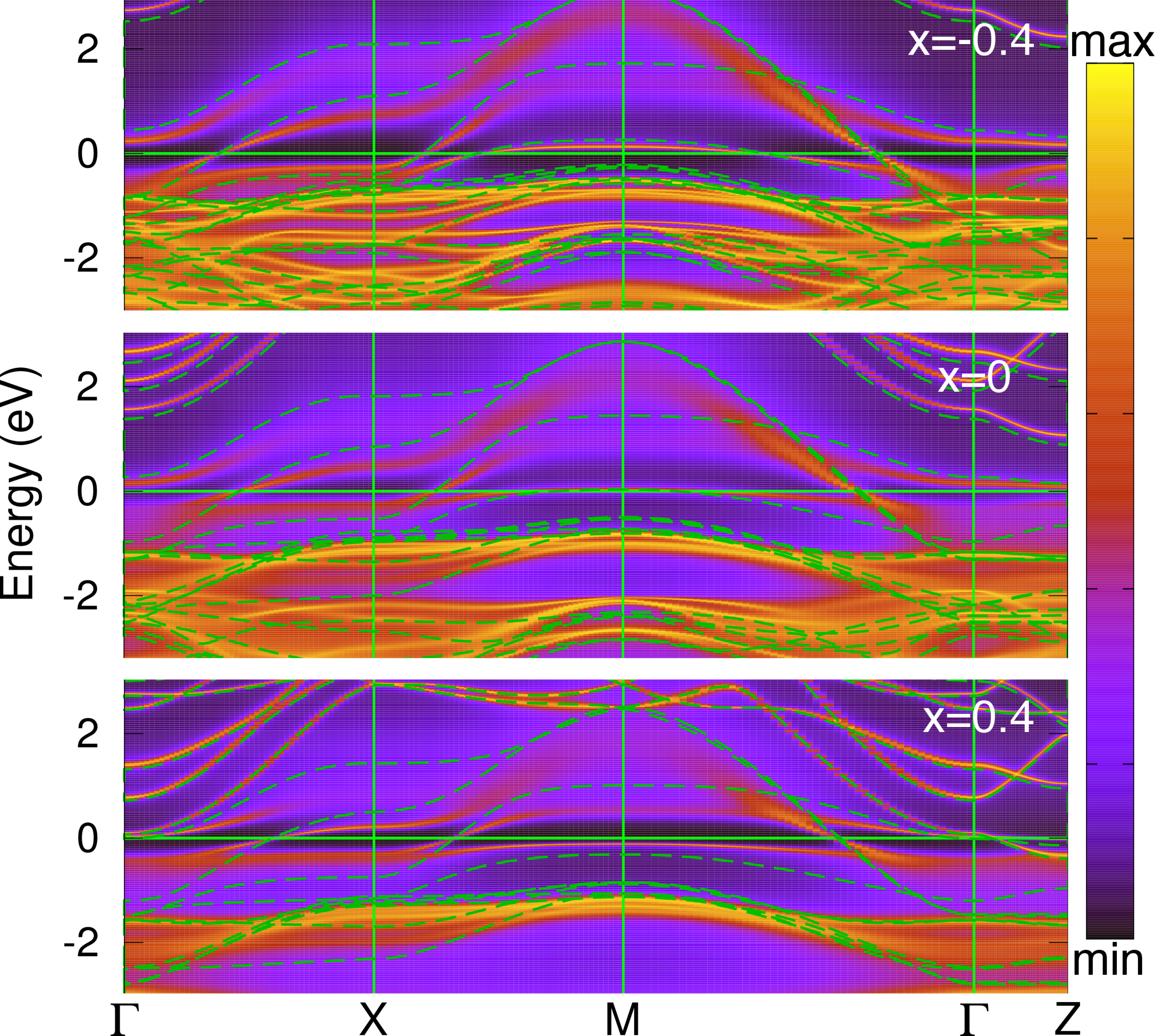}}
\caption{{\bf k}-resolved spectral functions of PM LNO as a function of doping obtained using DFT+DMFT with $U=6$ eV and $J=0.95$ eV at $T = 116$~K. The calculations are performed for the high-pressure orthorhombic crystal structure (determined experimentally at $\sim$30 GPa) with optimized atomic positions. We compare DFT+DMFT spectral functions with the nonmagnetic DFT results (shown with green broken lines).
}
\label{Fig_1}
\end{figure}

\begin{figure}[tbp!]
\centerline{\includegraphics[width=0.5\textwidth,clip=true]{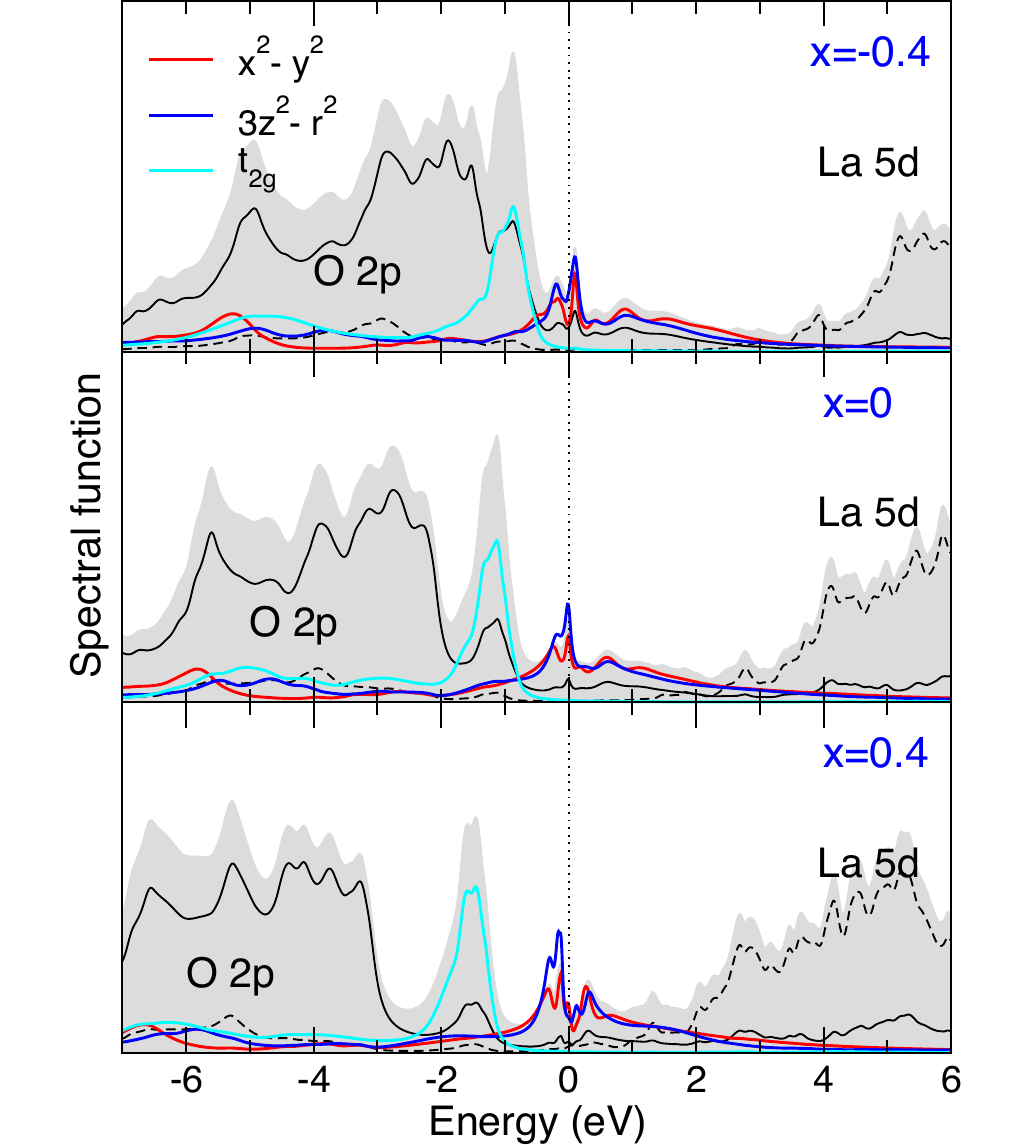}}
\caption{Orbital-dependent spectral functions of PM LNO as a function of doping obtained by DFT+DMFT at $T=116$K . The partial Ni $t_{2g}$, $x^2 - y^2$, and $3z^2 - r^2$ orbital contributions are shown (per orbital). The partial Ni $x^2 - y^2$, and $3z^2 - r^2$ orbital states are magnified by a factor 3 for better readability.
}
\label{Fig_2}
\end{figure}

In Fig.~\ref{Fig_1} we display our results for {\bf k}-resolved spectral functions of LNO obtained using DFT+DMFT for different levels of doping in comparison to the nonmagnetic DFT band structure. The {\bf k}-integrated orbital-dependent spectral functions are shown in Fig.~\ref{Fig_2}. Overall, our results agree well with previous electronic structure calculations of LNO \cite{Zhang_2023a,Shilenko_2023,Lechermann_2023,
Christiansson_2023,Liao_2023,Shen_2023b,Ryee_2024,
Lechermann_2024,Craco_2024,Cao_2024,Wu_2024,Leonov_2025}. We observe remarkable orbital-dependent bands renormalizations and strong incoherence of the spectral weights originating from the Ni $3d$ bands, caused by correlation effects. The partially occupied Ni $x^2-y^2$ and $3z^2-r^2$ states appear near the Fermi level. The occupied O $2p$ states are located at about -4 eV (for $x=0$) below the $E_F$, strongly hybridizing with the partially occupied Ni $3d$ orbitals. In agreement with previous calculations, the Wannier Ni $x^2-y^2$ and $3z^2-r^2$ orbital occupations are close to half-filling, of $\sim$0.54 and 0.6 per spin orbit (for $x=0$), respectively. The Ni $t_{2g}$ states are fully occupied and located at about -1 to -2 eV below $E_F$.

{Our results for} {\bf k}-resolved spectral functions evaluated as $A({\bf k},\omega) = -1/\pi~\mathrm{Im}[\omega + \mu - H_\mathrm{\bf k} - \Sigma( \omega )]^{-1}$ [where, $H_\mathrm{\bf k}$ is the self-consistent Kohn-Sham Hamiltonian
in the Wannier basis set, $\Sigma( \omega )$ is the self-energy
analytically continued on the real energy axis] show a remarkable bonding-antibonding splitting of the Ni $3z^2-r^2$ and $x^2-y^2$ bands near the Fermi level. 
{It is caused} by the strong inter-layer Ni-Ni coupling, which is characteristic for the double-layer LNO \cite{Zhang_2023a,Shilenko_2023,Lechermann_2023,
Christiansson_2023,Liao_2023,Shen_2023b,Ryee_2024,
Lechermann_2024}. 
{Note that in the bonding-antibonding molecular orbital basis the bonding and antibonding orbitals are defined as the $+$ and $-$ combinations of the Ni $e_g$ orbitals belonging to the top Ni$_A$ and bottom Ni$_B$ sites, such as $(3z^2-r^2)_\pm = (|3z^2-r^2\rangle_A \pm  |3z^2-r^2\rangle_B)/\sqrt{2}$.}
The bonding Ni {$(3z^2-r^2)_+$} states are nearly fully occupied, form a shallow flat quasiparticle band crossing the Fermi level near the Brillouin zone (BZ) M-point (for $x=0$). {For the nominally Ni$^{2.5+}$ ion} this implies a nearly quarter filling of the planar Ni {$(x^2-y^2)_\pm$} orbitals. 
{In a real system, however, this simplified picture is further complicated by charge transfer between the Ni $3d$ and O $2p$ states.}
Our results for $A({\bf k},\omega)$ reveal nearly flat band dispersions at about 60~meV below and 0.4 eV above $E_F$ at the X-point on the BZ $\Gamma$-X branch, originating from the bonding-antibonding splitted Ni $x^2-y^2$ orbitals. This suggests a van Hove anomaly in the electronic spectra of LNO near the Fermi level, which is associated with the quasi-2D nature of its electronic structure, in close similarity to cuprates.


\begin{table}[h]
{
\centering
\caption{DFT+DMFT results for doping dependence of the total Ni $3d$, orbitally-resolved Ni $x^2 - y^2$ and $3z^2 - r^2$ Wannier occupations and the corresponding bonding Ni $e_g$ [$(x^2 - y^2)_-$ and $(3z^2 - r^2)_+$] states (per spin orbit).}
\begin{ruledtabular}
\begin{tabular}{cccccccc}

\multicolumn{1}{c}{Doping (el./Ni) } & \multicolumn{1}{c}{Ni $3d$} &  \multicolumn{4}{c}{Orbital occupations}  \\
& & $x^2-y^2$ & $3z^2-r^2$ & $(x^2 - y^2)_-$ & $(3z^2 - r^2)_+$ \\

\hline
-0.4 &  8.13 & 0.52 & 0.57  & 0.52 & 0.66 \\ 
-0.3 &  8.16 & 0.53 & 0.58  & 0.52 & 0.68 \\
-0.2 &  8.18 & 0.53 & 0.58  & 0.54 & 0.69 \\
-0.1 &  8.21 & 0.54 & 0.59  & 0.55 & 0.70 \\ 
   0 &  8.24 &  0.54 & 0.60    & 0.57 & 0.72 \\ 
0.1 &  8.27 &  0.55 & 0.61  & 0.58 & 0.73 \\ 
0.2 &  8.30 &  0.56 & 0.62  & 0.59 & 0.76 \\ 
0.3 &  8.34 &  0.57 & 0.64  & 0.60   & 0.79 \\
0.4 &  8.37 &  0.58 & 0.65  & 0.61 & 0.80 \\ 
\end{tabular}
\end{ruledtabular}
\label{tab_1}}
\end{table}

\begin{figure}[tbp!]
\centerline{\includegraphics[width=0.5\textwidth,clip=true]{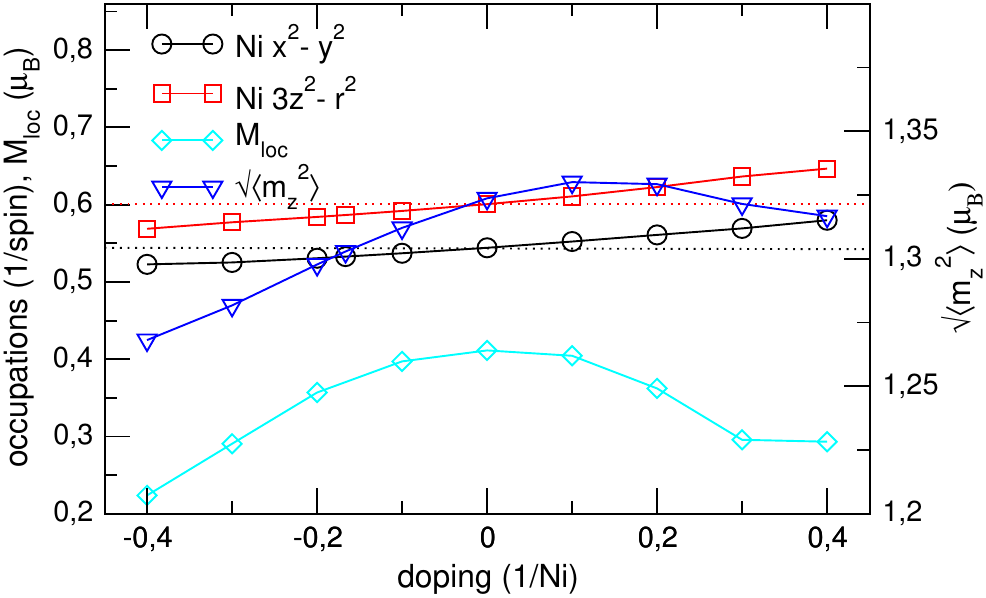}}
\caption{Ni $e_g$ orbital occupations, instantaneous magnetic moments ($\sqrt{\hat{m}^2_z}$) and fluctuating local moments ($M_\mathrm{loc}$) as a function of doping calculated by DFT+DMFT with $U=6$ eV and $J=0.95$ eV at $T=116$~K. The fluctuating magnetic moments evaluated as $M_\mathrm{loc}=[k_BT \int \chi(\tau) d\tau ]^{1/2}$), where $\chi(\tau) \equiv \langle \hat{m}_z(\tau)\hat{m}_z(0) \rangle$ is the local spin-spin correlation function.
}
\label{Fig_3}
\end{figure}

\begin{figure}[tbp!]
\centerline{\includegraphics[width=0.5\textwidth,clip=true]{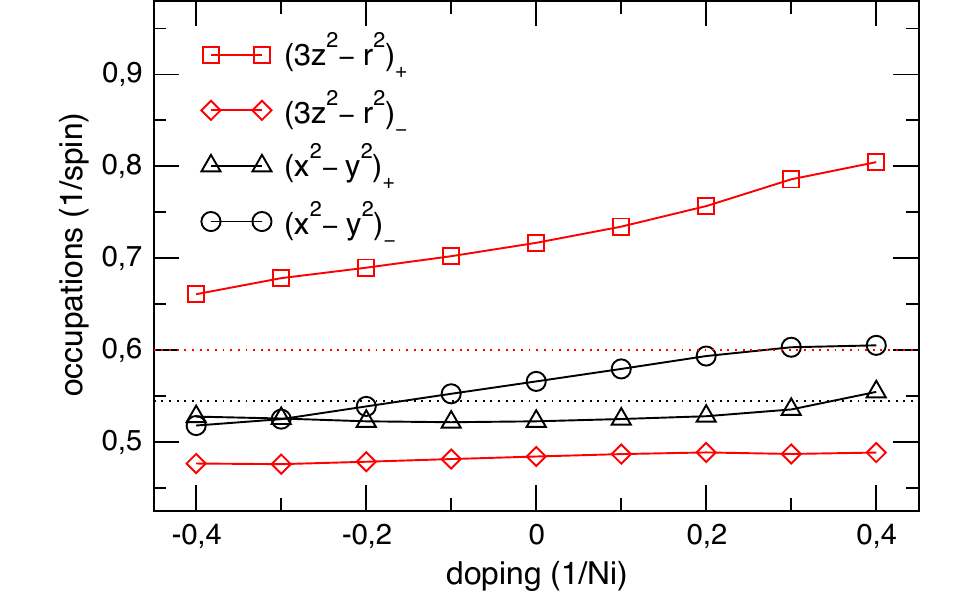}}
\caption{
{
Ni $3z^2-r^2$ and $x^2-y^2$ orbital occupations in the bonding-antibonding molecular orbital basis as a function of doping calculated by DFT+DMFT with $U=6$ eV and $J=0.95$ eV at $T=116$~K. }
}
\label{Fig_4}
\end{figure}

{Our results for doping dependence of Ni $3d$ orbital occupations are summarized in Table~\ref{tab_1}.} We find that doping results in a nearly linear change of the calculated Ni $x^2-y^2$ and $3z^2-r^2$ Wannier orbital occupations as shown in Fig.~\ref{Fig_3}. 
{In agreement with this, we observe a nearly linear increase in the orbital occupancy of the bonding $(3z^2-r^2)_+$ and $(x^2-y^2)_-$ states (see Fig.~\ref{Fig_4}). The occupations of the antibonding Ni $e_g$ orbitals remain nearly constant with doping.}
Our DFT+DMFT calculations exhibit significant redistribution of the quasiparticle spectral weights upon doping. It is accompanied by a remarkable shift by $\sim$ 1.4 eV (with respect to that at $x=0$) of the occupied O $2p$ bands towards the Fermi level at $x=-0.4$, which leads to a sharp increase of hybridization between the Ni $3d$ and O $2p$ states. This leads to a significant charge transfer between the Ni $e_g$ and O $2p$ states, which implies a crossover in the negative charge-transfer state \cite{Zaanen_1985}.

Upon electron doping the effect is opposite. It is accompanied by an entire change of the Ni  $x^2-y^2$ and $3z^2-r^2$ orbital spectra near $E_F$, with a pseudogap like behavior upon electron doping to $x=0.4$. In addition, we observe a large shift of the unoccupied La $5d$ bands toward the $E_F$. For $x>0.2$ it leads to a crossover to a self-doping regime which is characterized by partial occupation of the La $5d$ orbitals. In fact, 
the infinite-layer nickelates $R$NiO$_2$ show similar behavior of the electronic structure with the rare-earth $5d$ bands crossing
the Fermi level \cite{Lee_2004,Kitatani_2020,Chen_2022a,Nomura_2022,
Botana_2022,Werner_2020,Karp_2020a,Lechermann_2020b,
Wang_2020,Leonov_2020b,Leonov_2021,Lechermann_2022}.
It is interesting that doping effects are hardly affect the position of the occupied Ni $x^2-y^2$ van Hove singularity, which is pined near to -60~meV for a broad range of $x$ between $x=-0.4$ and 0.4. At the same time, the position of the unoccupied (antibonding)  Ni $x^2-y^2$ van Hove state is strongly affected by doping. For instance, it shifts to about 40~meV above the $E_F$ upon elecctron doping x=0.4 (see Fig.~\ref{Fig_1}).

Our results for the orbital-dependent local spin susceptibility $\chi(\tau) \equiv \langle \hat{m}_z(\tau) \hat{m}_z(0) \rangle$
suggest that magnetic correlations in LNO are at
the verge of the formation of local magnetic
moments.
Thus, the calculated instantaneous magnetic
moment of Ni for $x=0$ is about 1.3~$\mu_\mathrm{B}$,  consistent with a nearly $S = 1/2$ state of Ni. At the same time, the fluctuating moment evaluated within DMFT as $M_\mathrm{loc} \equiv [k_\mathrm{B}T \int \chi(\tau)d\tau]^{1/2}$
is significantly smaller, $\sim$0.4~$\mu_B$ (for $x=0$).
We find a remarkable decrease of both the instantaneous and fluctuating moments upon doping (see Fig.~\ref{Fig_3}). We note however that the maximum of the instantaneous
moment  $\sqrt{\hat{m}_z^2}$ appears at about $x=0.1$, while the overall trend upon doping is consistent with that for $M_\mathrm{loc}$. 

\begin{figure}[tbp!]
\centerline{\includegraphics[width=0.5\textwidth,clip=true]{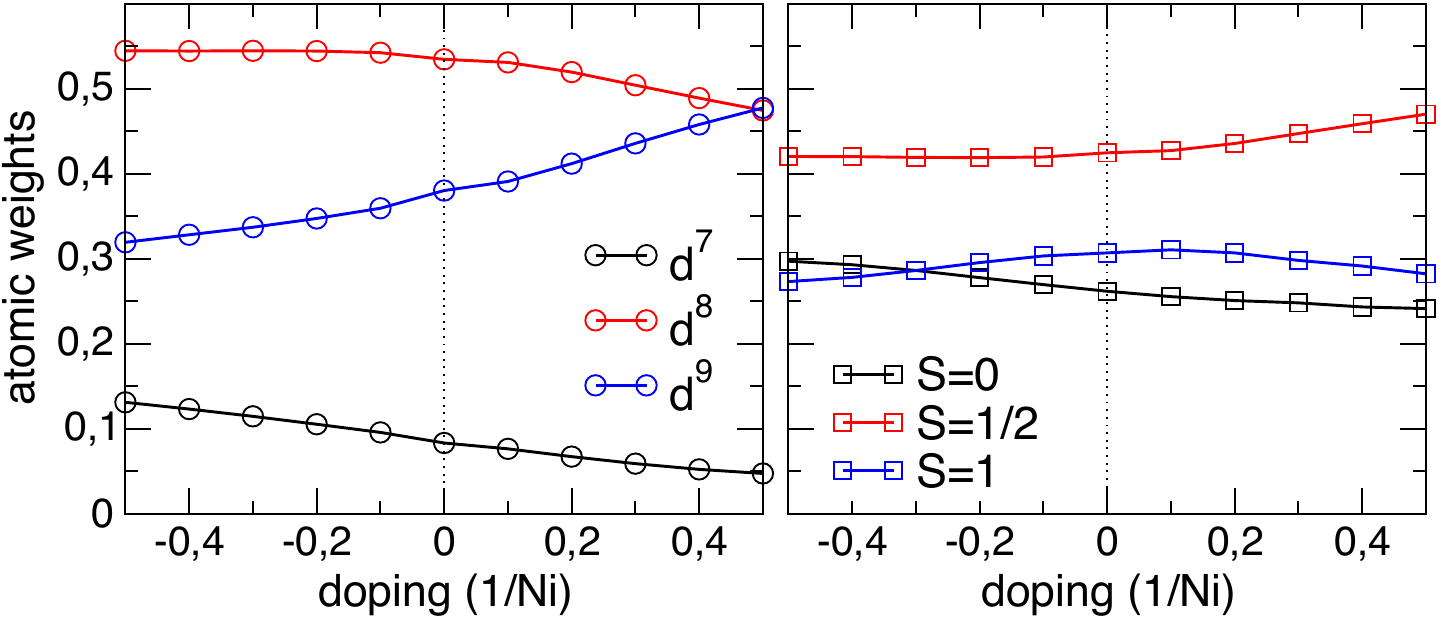}}
\caption{Our results for the weights of differnt atomic configururations (left panel: for the Ni $3d^7$, $3d^8$ and $3d^9$ valence configurations; right panel: for the spin-state $S=0$, $S=1/2$ and $S=1$ configurations) of the Ni $3d$ electrons calculated using DFT+DMFT for PM LNO at $T = 116$~K as a function of doping.
}
\label{Fig_5}
\end{figure}

These findings agree with 
our analysis of the weights of different atomic configurations of the Ni $3d$ electrons (in DMFT the Ni $3d$ states are seen fluctuating between various atomic configurations) shown in Fig.~\ref{Fig_5}. For a broad range of doping, we find a strong mixture of the Ni $3d^8$ and $3d^9$ (i.e., a $3d^9\underline{L}$ state with a hole localized in the O $2p$ band) valence configurations of the Ni ions, with a major contribution from the Ni$^{2+}$ ($3d^8$) valence configuration. Upon high electron doping, the electronic state is characterized by a nearly equal contribution of the Ni$^{2+}$ and Ni$^{+}$ valence states (a mixed-valence state of the Ni ions). Our analysis suggests that LNO appears close to a (negative) charge transfer regime, implying the importance of charge transfer effects upon doping. Moreover, our estimate for the spin-state configuration gives a predominant contribution of the $S=1/2$ state. This state is accompanied by strongly competing the $S=0$ and $S=1$ states in the electronic structure of doped LNO.

\subsection{Fermi surface }

\begin{figure}[tbp!]
\centerline{\includegraphics[width=0.5\textwidth,clip=true]{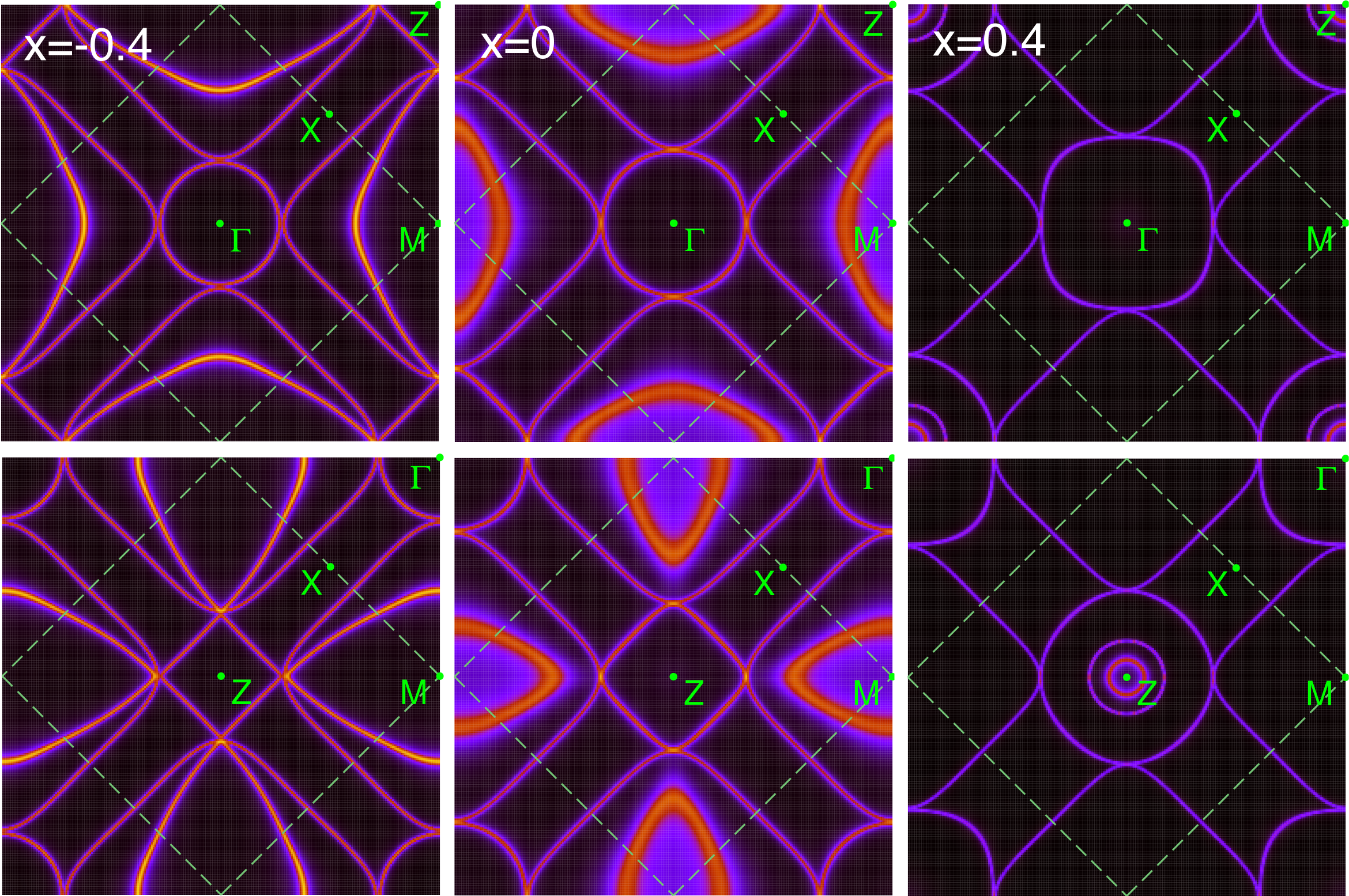}}
\caption{Quasiparticle Fermi surfaces [spectral function $A( {\bf k}, \omega)$ evaluated at $\omega = 0$] for ${\bf k}_z=0$ (top panel) and ${\bf k}_z=\pi/c$ (bottom panel) calculated using DFT+DMFT for PM LNO at $T = 116$~K for different doping levels.
}
\label{Fig_6}
\end{figure}

Our DFT+DMFT results exhibit a remarkable reconstruction of the low-energy electronic states, accompanied by an entire change of the Fermi surface (FS) upon doping, i.e., a Lifshitz transition. In Fig.~\ref{Fig_6} we summarize our results for the quasiparticle FS evaluated within DFT+DMFT for different doping. In agreement with previous results for $x=0$ \cite{Shilenko_2023,
Lechermann_2023,Christiansson_2023,Liao_2023,Shen_2023b,
Ryee_2024,Lechermann_2024}, we obtain a quasi-2D FS, dominated by a large electronlike FS ($\alpha$-sheet) centered at the $\Gamma$-point and large holelike FSs ($\beta$-sheet) near the M-point, predominantly of the Ni $x^2 - y^2$ orbital character, closely resembling the FS of the optimally hole-doped cuprates. In addition, the calculations show the holelike FS ($\gamma$-sheet) around the M-point, originating from the nearly completely occupied bonding Ni {$(3z^2-r^2)_+$} orbitals, highlighting the multi-orbital nature of the low-energy excitations in LNO, in contrast to cuprates.

We observe a large enhancement of the holelike $\gamma$-pocket upon hole doping. In contrast to this, the electron doping tends to suppress the $\gamma$-pocket, obviously due to the full occupation of the bonding Ni {$(3z^2-r^2)_+$} quasiparticle bands which are entirely shifted below $E_F$ for $x>0.2$. This leads to reconstructions of the low-energy electronic structure of LNO. Interestingly that in a narrow region of electron doping $0.2< x< 0.3$ we observe a crossover toward the two-sheets FSs (consisting of the $\alpha$ and $\beta$ FS pockets) predominantly of the Ni $x^2 - y^2$ orbital character. That is, the $\gamma$-sheet around the M-point disappears, resulting in a nominally single-orbital Ni $x^2 - y^2$ character of the low-energy excitations in the electron-doped LNO. 

In addition, we find a large shift of the unoccupied antibonding Ni {$(3z^2-r^2)_-$}  and La $5d$ states toward the $E_F$, seen on the BZ $\Gamma$-Z and $\Gamma$-X. Most importantly, upon electron doping above $x>0.3$, these quasiparticle bands cross the Fermi level, i.e., are partially occupied, resulting in the emergence of the quasi-3D FS pocket(s) (presumably of the electronlike character) centered near the BZ $Z$-point. This suggests the importance of the effects of electron-electron correlations and self-doping to explain the electronic structure of the electron-doped LNO. Moreover, we notice a remarkable variation with doping of the shape of the central $\alpha$-sheet of the Ni $x^2-y^2$ orbital character near the $\Gamma$-point, from a circlelike for $x<0.1$ to a squarelike shape for $x>0.2$ for ${\bf k}_z = 0$ (this behavior is in fact opposite for ${\bf k}_z = \pi/c$), with a large increase of its size.

\subsection{Orbital-selective renormalizations}

\begin{figure}[tbp!]
\centerline{\includegraphics[width=0.5\textwidth,clip=true]{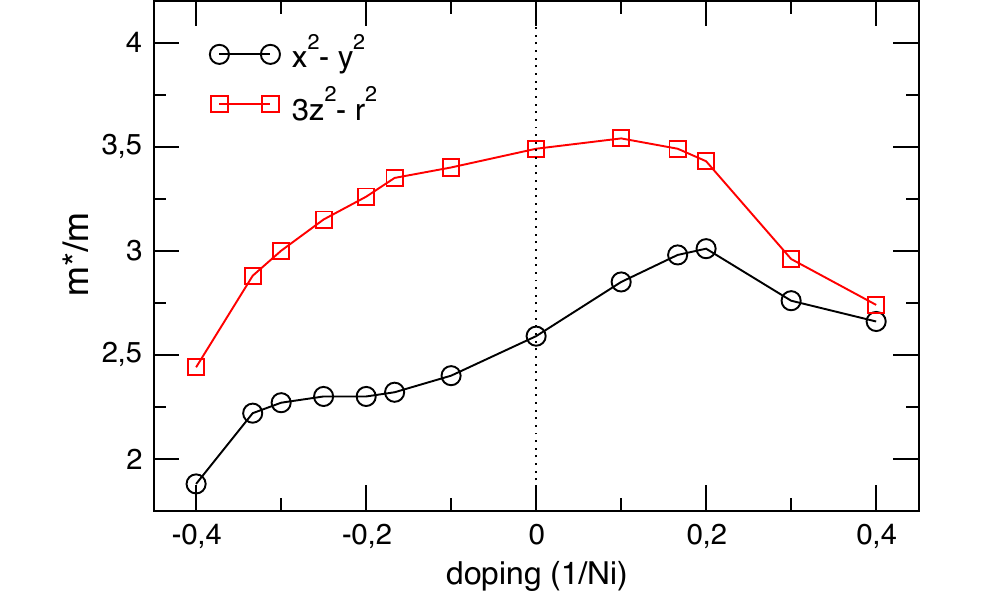}}
\caption{Orbital-dependent quasiparticle mass renormalizations  $m^*/m$ for the Ni $x^2-y^2$ and $3z^2-r^2$ orbitals of PM LNO as a function of doping calculated by DFT+DMFT with $U=6$ eV and $J=0.95$ eV at $T=116$~K. 
}
\label{Fig_7}
\end{figure}

Our analysis of the DFT+DMFT results show significant orbital-selective renormalizations of the Ni $x^2 - y^2$ and $3z^2 - r^2$ bands (as shown in Fig.~\ref{Fig_7}; see also Table~\ref{tab_2}). This implies orbital-dependent localization of the Ni $3d$ electrons. Moreover, the calculated Ni $3d$ self-energies show a Fermi liquid-like behavior, with a substantial orbital-dependent quasiparticle damping of $Im[\Sigma(i\omega_n)] \sim 0.1$ and 0.14 eV for the Ni $x^2-y^2$ and $3z^2-r^2$ states at the first Matsubara frequency, at $T = 116$ K. 
The fully occupied Ni $t_{2g}$ states are sufficiently coherent, with $Im[\Sigma(i\omega_n)]$ below 0.01~eV at the first Matsubara frequency. In Fig.~\ref{Fig_7} we show our results for the orbital-dependent quasiparticle mass renormalizations $m^*/m$, caused by strong correlations of the Ni $3d$ electrons for different doping levels. 
We use $\frac{m^*}{m} = [1 - \partial \mathrm{Im}[\Sigma(i\omega)]/\partial i\omega]|_{i\omega\rightarrow 0}$ to evaluate $m^*/m$, fitting the self-energy results $\Sigma(i\omega)$ at the lowest eight Matsubara frequencies for the Ni $3d$ bands. Pad\'e approximants give similar results. 

In agreement with previous results for the undoped LNO \cite{Shilenko_2023,Lechermann_2023,Christiansson_2023,Liao_2023,Shen_2023b,
Ryee_2024,Lechermann_2024}, the Ni $3z^2-r^2$ orbitals are seen to be more correlated and incoherentlike than the planar $x^2-y^2$ orbitals. This result is consistent with a more narrow DFT bandwidth of the Ni $3z^2-r^2$ orbitals which is by $\sim$19\% less than that for the $x^2-y^2$ states for $x=0$. The effective mass renormalization of the Ni $t_{2g}$ orbitals is much weaker and nearly doping independent, of $\sim$1.2. Upon doping, we obtain a large variation of the orbital-dependent quasiparticle band renormalizations. In fact, the $3z^2-r^2$ orbitals show a nearly flat doping-independent behavior upon a moderate doping from $x=-0.2$ to 0.2, followed by a large drop from $m^*/m \sim 3.5$ to 2.5, by about 30\%, upon further doping to $x=-0.4$ and 0.4. The Ni $x^2-y^2$ bands exhibit large variations with two well-defined maxima at about $x=-0.3$ and $x=0.2$, which are presumably associated with the Lifshitz transitions. Upon electron doping above $x=0.2$, the Lifshitz transition is accompanied by a change of the low-energy electronic structure, associated with self-doping. This leads to suppression of the effects of electronic correlations, seen as a large decrease of $m^*/m$ for the Ni $x^2-y^2$ and $3z^2-r^2$ orbitals. For $x \sim 0.4$ the quasiparticle renormalizations are seen to be nearly orbital-independent (as expected for the nominal Ni$^{2+}$ state).

\begin{table}[h]
{
\centering
\caption{Ni $3d$, Ni $x^2 - y^2$ and $3z^2 - r^2$ Wannier occupations, and orbital-dependent quasiparticle mass enhancement $m^*/m$ of the Ni $e_g$ states as a function of doping obtained by DFT+DMFT with $U=6$ eV and $J=0.95$ eV at $T=116$~K.}
\begin{ruledtabular}
\begin{tabular}{cccccccc}

\multicolumn{1}{c}{Doping (el./Ni) } & \multicolumn{1}{c}{Ni $3d$} &  \multicolumn{2}{c}{Orbital occupations} & \multicolumn{2}{c}{$m^*/m$} \\
& & $x^2-y^2$ & $3z^2-r^2$ & $x^2-y^2$ & $3z^2-r^2$ \\

\hline
-0.4 &  8.13 & 0.52 & 0.57  & 1.88 & 2.44 \\ 
-0.3 &  8.16 & 0.53 & 0.58  & 2.27 & 3.0 \\
-0.2 &  8.18 & 0.53 & 0.58  & 2.3 & 3.26 \\
-0.1 &  8.21 & 0.54 & 0.59  & 2.4 & 3.4 \\ 
   0 &  8.24 &  0.54 & 0.60    & 2.59 & 3.49 \\ 
0.1 &  8.27 &  0.55 & 0.61  & 2.85 & 3.54 \\ 
0.2 &  8.30 &  0.56 & 0.62  & 3.01 & 3.43 \\ 
0.3 &  8.34 &  0.57 & 0.64  & 2.76   & 2.96 \\
0.4 &  8.37 &  0.58 & 0.65  & 2.66 & 2.74 \\ 

\end{tabular}
\end{ruledtabular}
\label{tab_2}}
\end{table}

\subsection{Magnetic correlations}

\begin{figure}[tbp!]
\centerline{\includegraphics[width=0.5\textwidth,clip=true]{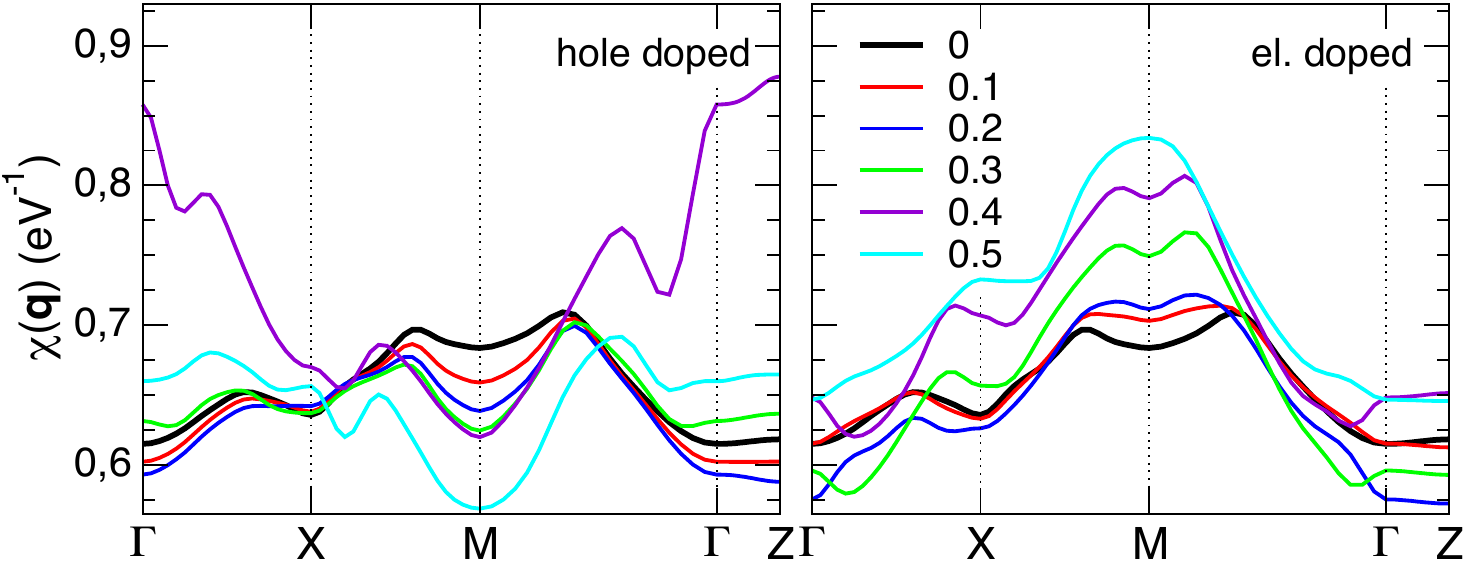}}
\caption{Our results for static spin susceptibility $\chi({\bf q})$ of PM LNO as a function of doping calculated within DFT+DMFT using the particle-hole bubble approximation (neglecting the high-order vertex corrections) at $T=116$~K. Our results for the electron (right panel) and hole dopings (left panel) ranging from the nominal Ni$^{2+}$ (electron doping $x=0.5$) to Ni$^{3+}$ (hole doping $x=-0.5$) state of the double-layer LNO are shown.
}
\label{Fig_8}
\end{figure}

Next, we perform analysis of the strength of magnetic correlations in LNO, associated with multiple in-plane nesting of the Fermi surface. Using the particle-hole bubble approximation within DFT+DMFT (neglecting the high-order vertex corrections), we calculate the momentum-dependent static magnetic susceptibility as $\chi({\bf q})=-k_BT~\mathrm{Tr}[\Sigma_{{\bf k}, i\omega_n}G_{\bf k}(i\omega_n)G_{{\bf k}+{\bf q}}(i\omega_n)]$. In this expression, $G_{\bf k}(i\omega_n)$ is the local interacting Green's function for the Ni $3d$ states evaluated on the Matsubara contour $i\omega_n$ within DFT+DMFT. In Fig.~\ref{Fig_8} we display our results for $\chi({\bf q})$ as a function of the electron (right panel) and hole dopings (left panel) ranging from the nominal Ni$^{2+}$ (electron doping $x=0.5$) to Ni$^{3+}$ (hole doping $x=-0.5$) state of the double-layer LNO. We find that upon high electron doping $x=0.5$, i.e., for the nominal Ni$^{2+}$ state, $\chi({\bf q})$ exhibits a well defined maximum with a commensurate wave vector at the M-point, which is accompanied by a concomitant substructure at the X-point. This result suggests the N\'eel ($G$-type) AFM ordering for the $x=0.5$ electron-doped LNO (with the Ni$^{2+}$ ions). This result is in qualitative agreement with the behavior of the structurally-related single-layer RP La$_2$NiO$_{4+\delta}$ with a nominal Ni$^{2+}$ state \cite{Yamada_1992,Fabbris_2017,Petsch_2023}. In fact, nearly to its stoichiometric composition La$_2$NiO$_{4+\delta}$ exhibits the N\'eel-type AFM state accompanied by weak ferromagnetism at low temperatures, arising from canting of the Ni $3d$ moments.

For a broad range of dopings from $x=0.4$ to -0.4, we observe the appearance of two asymmetric maxima of $\chi({\bf q})$ near the M-point, characterized by two incommensurate wave vectors located on the BZ $\Gamma$-M and X-M branches.  It is accompanied by a minor instability with a propagating vector at the $\Gamma$-X, which is in qualitative agreement with the behavior of $\chi({\bf q})$ found previously for LNO ($x=0$). This result indicates the suppression upon hole doping of the N\'eel-type long-range AFM ordering of the Ni$^{2+}$ ions. The latter is seen as the appearance of a pronounced deep in $\chi({\bf q})$ at the M-point for all $x<0.5$. As a result, this leads to the formation of competing spin- and change-density wave states in LNO. We therefore propose the emergence of spin-charge-density-wave stripes in LNO upon doping, characterized by the in-plane long-range ordering of the Ni $3d$ moments in the basal $ab$ in-plane of the NiO$_6$ bilayer (with a propagating wave vector oriented at 45$^0$ to the Ni-O bond). We note that this behavior is in close similarity to previously discussed the double spin-charge-density-wave stripe ordering in the double-layer LNO (with $x=0$) at low pressure and temperature \cite{Chen_2024a,Kakoi_2024,Agrestini_2024,Dan_2024,
Leonov_2025,LaBollita_2024b,BZhang_2024,Ni_2024,Tian_2025}. The latter is characterized by a propagating wave vector $\mathrm{\bf q}=(\frac{1}{4},\frac{1}{4})$ arrangement (oriented at 45$^0$ to the Ni-O bond) of the nominally high-spin Ni$^{2+}$ and low-spin Ni$^{3+}$ (with a predominant $3d^8\underline{L}$ configuration with a hole localized in the O $2p$ band) ions which form zigzag ferromagnetic chains alternating in the $ab$ plane \cite{Leonov_2025}.

Upon further hole doping near to $x=-0.4$, we observe a remarkable reconstruction of magnetic correlations, accompanied with the appearance of the commensurate A-type AFM state. The latter is followed by a competing spin and change density wave state at $x=-0.5$ (for the nominal Ni$^{3+}$ state), characterized by two incommensurate wave vectors on the $\Gamma$-X and $\Gamma$-M branches. This is suggestive that a stripe state propagating along the Ni-O bond [associated with an instability in $\chi({\bf q})$ at the BZ $\Gamma$-X branch] appears to play a role. We link this reconstruction of magnetic correlations with the Lifshitz transition of the electronic structure of LNO near $x=-0.4$, which is accompanied by a change of the Fermi surface topology.

Our analysis of the static magnetic susceptibility $\chi({\bf q})$  suggests the possible formation of the spin and charge (or bond) density wave stripes caused by nesting of the Fermi surface of LNO. It is  complicated by strong superexchange interactions between the Ni ions, implying strong spin and charge correlations in LNO. We observe that spin and charge stripe fluctuations play a key role in tuning pressure-driven superconductivity and are strongly affected by doping (stoichiometry) of LNO. Indeed, we find a significant enhancement of the strength of in-plane spin and charge fluctuations upon a moderate electron doping of the Ni$^{2.5+}$ ions (e.g., upon oxygen deficiency). This behavior is linked to the Lifshitz transition, associated with the disappearance of the holelike FS $\gamma$-sheet (originating from the bonding Ni $3z^2-r^2$ bands) of LNO. We note a close resembles of the obtained results to those for the bilayer Hubbard model, in which superconductivity is found to be enhanced as one of the two electron bands approaches the Lifshitz transition and, in particular, when it becomes incipient (e.g., due to doping) \cite{Karakuzu_2021}. Our results suggest that spin and charge stripe fluctuations, effectively tuned by doping, play a key role in  pressure-driven superconductivity of LNO.

\section{Conclusion}

In conclusion, we presented a comprehensive DFT+DMFT study of the electronic structure and magnetic properties of highly-pressurized La$_3$Ni$_2$O$_{7-\delta}$ upon doping.  
Our results show a sensitive dependence of orbital-dependent band renormalizations upon doping, implying the proximity of the Ni $3d$ states to orbital-dependent localization.
We found a remarkable reconstruction of the low-energy electronic structure of LNO upon doping, accompanied by the Lifshitz transition(s). It is accompanied by a crossover to the self-doping regime characterized by partial occupation of the La $5d$ bands (upon electron doping $x>0.2$), in close similarity to the infinite-layer systems.
Our analysis suggests possible formation of the spin and charge (or bond) density wave stripes, implying strong spin and charge correlations in LNO. 
We show that this behavor is associated with suppression of the N\'eel AFM ordering of the Ni$^{2+}$ ions upon hole doping. 
Most importantly, upon a moderate electron doping of the Ni$^{2.5+}$ ions, e.g., upon oxygen deficiency, we obtain a significant enhancement the strength of magnetic correlations, associated with the Lifshitz transition. As a result, we expect a sharp increase of planar spin- and charge-density-wave fluctuations, which are sensitively tuned by doping. {We therefore expect that electron doping, e.g., upon doping LNO with Ce, can enhance a superconducting transition temperature in LNO.}
{This result closely resemble that} for the bilayer Hubbard model, which show the boosting of superconductivity as one of the two electron bands approaches the Lifshitz transition and, in particular, when it becomes incipient (e.g., due to doping). 
Our results suggest that spin and charge stripe fluctuations, effectively tuned by doping, play a key role in pressure-driven superconductivity in LNO.

\section{ACKNOWLEDGMENTS}
We thank V. I. Anisimov for helpful discussions. The DFT electronic structure calculations were supported within the framework of the state assignment of the Ministry of Science and Higher Education of the Russian Federation for the IMP UB RAS. The DFT+DMFT calculations, theoretical analysis of the electronic structure and magnetic properties were supported by the Russian Science Foundation (Project No. 25-12-00416).

\end{document}